\documentclass[10pt]{article}
\newcommand{\beq}{\begin{equation}}
\newcommand{\eeq}{\end{equation}}
\newcommand{\beqn}{\begin{eqnarray}}
\newcommand{\eeqn}[1]{\label{#1}\end{eqnarray}}

\newcommand{\id}
{i\kern.06em\hbox{\raise.25ex\hbox{$/$}\kern-.60em$\partial$}}

\newcommand{\bs}{/\kern-.52em b}
\newcommand{\qs}{/\kern-.52em s}

\newcommand{\dd}
{\kern.06em\hbox{\raise.25ex\hbox{$/$}\kern-.60em$\partial$}}

\newcommand{\grad}{\vec\bigtriangledown}
\begin{document}

\begin{titlepage}

\title{On the Renormalization of Hamiltonians}

\author{
G. Alexanian$^a$\thanks{E-mail address:
garnik@scisun.sci.ccny.cuny.edu}\,\,\,\,\, and \,\,\,\,\,
E. F. Moreno$^{a,b}$\thanks{E-mail address:
moreno@scisun.sci.ccny.cuny.edu}\\
\\
~
\\
{\small\it $^a$ Physics Department,
City College of the City University of New York}\\
{\small\it New York NY 10031, USA}\\
~\\
{\small\it $^b$ Baruch College,
City University of New York}\\
{\small\it New York NY 10010, USA}\\
}

%\date{}

\maketitle

%\maketitle

\begin{abstract}
We introduce a novel method for the renormalization of the
Hamiltonian operator in Quantum Field Theory in the spirit of the
Wilson renormalization group. By a series of unitary
transformations that successively  decouples the high-frequency
degrees of freedom and partially diagonalizes the high-energy
part, we obtain the effective Hamiltonian for the low energy
degrees of freedom. We successfully apply this technique to
compute the 2-loop renormalized Hamiltonian in scalar $\lambda\;
\phi^4$ theory.
\end{abstract}
\end{titlepage}

%%%%%%%%%%%%%%%%%%%%%%%%%%%%%%%%%%%%%%%%%%%%%%%%%%%%%%%%%%%%%%%%%%%%%%%%
%%%%%%%%%%%%%%%%%%%%%%%%%%%%%%%%%%%%%%%%%%%%%%%%%%%%%%%%%%%%%%%%%%%%%%%%
%%%%%%%%%%%%%%%%%%%%%%%%%%%   SECTION  I  %%%%%%%%%%%%%%%%%%%%%%%%%%%%%%
%%%%%%%%%%%%%%%%%%%%%%%%%%%%%%%%%%%%%%%%%%%%%%%%%%%%%%%%%%%%%%%%%%%%%%%%
%%%%%%%%%%%%%%%%%%%%%%%%%%%%%%%%%%%%%%%%%%%%%%%%%%%%%%%%%%%%%%%%%%%%%%%%

\section{Introduction}
The subject of renormalization for the Hamiltonian operator has
been looked into extensively for the past decade \cite{GlWi}. Yet
it is hardly being used in any practical computation in field
theory, since usually all the intermediate steps are
non-covariant and, generally, rather complicated compared to the
standard approach. Nevertheless the concept of ``integrating
out'' of the high frequency degrees of freedom has much more
physical meaning precisely in the Hamiltonian framework. One way
to present it would be to set up the whole procedure as a
Born-Oppenheimer approximation used in atomic physics
\cite{Nair-Minic}. Another way to look at it would be to
introduce a unitary transformation in order to decouple ``high''
and ``low'' modes and then look at the low-energy part of the
spectrum \cite{We}. In this letter we want to introduce a
renormalization technique in the spirit of the Wilsonian
renormalization \cite{wilson}, {\it i.e.,} integration of fast
degrees of freedom, appropriate for the Hamiltonian
(Schr\"odinger) formalism \cite{j}. It is similar in essence to
both \cite{Nair-Minic} and \cite{We}, but appears to be very
different in practice. Basically what we want to see is how the
high frequency degrees of freedom modify the low energy
Hamiltonian operator, or equivalently, what Hamiltonian in term
of low frequency modes produces the same low energy physics as
the original Hamiltonian. This technique was successfully applied
in a previous paper by the authors to the 1-loop renormalization
of Abelian and non-Abelian gauges theories \cite{AM}.

For the sake of explicitness let us consider a system described by a
Hamiltonian $H(\Lambda)$ defined up to the momentum scale
(cut-off) $\Lambda$; that is, $H(\Lambda)$ produces finite results.
Now if $\mu$ is a lower momentum scale, we want to find a new
Hamiltonian operator $H(\mu)$ that generates the same results for
all the physical processes which do not involve momenta greater than
$\mu$. The definition of $H(\mu)$ is simple: $H(\mu)$ is the
projection of the original Hamiltonian $H(\Lambda)$ onto the low
frequency subspace, {\it i.e.,} the high frequency vacuum,
\begin{equation}
H(\mu) = {\cal P}_{low}\; H(\Lambda)\; {\cal P}_{low} .
\label{hr1}
\end{equation}
The claim is that, at least perturbatively, it is possible to
decouple the low frequency modes from the high frequency modes
and thus give sense to the notion of ``low frequency subspace".
In essence we will show that after a suitable unitary
transformation we can partially diagonalize the Hamiltonian and
construct the vacuum for high momentum modes. Therefore we can
rewrite eq.~(\ref{hr1}) as
\begin{equation}
H(\mu) = \langle 0_{high} |U^{\dagger}(\Lambda,\mu)\;
           H(\Lambda)\;U(\Lambda,\mu)| 0_{high}\rangle.
\label{hr2}
\end{equation}

Now let us explain how to construct the unitary operator
$U(\Lambda,\mu)$. As we have explained before, the main goal is
to identify the ground state of the high frequency modes and then
project the whole Hamiltonian onto this state. So if we can find
a unitary transformation that: {\it i)} separates the low energy
modes from the high energy modes and, {\it ii)} diagonalizes the
high energy part of the Hamiltonian (for example in terms of the
creation and annihilation operators of the high frequencies), we
are done: the ground state will be then the state annihilated by
all the high frequency annihilation operators. This task is of
course difficult, but in fact we need less than that. Since we
only need to identify the vacuum state of the high momenta
Hamiltonian and not the whole spectrum, only a partial
diagonalization is enough. In fact, only those non-diagonal terms
containing purely creation operators or purely annihilation
operators have to be removed from the Hamiltonian. Other
non-diagonal terms containing both creation and annihilation
operators do not change the vacuum state of the theory (once the
aforementioned terms have been removed).

We now show how this work in more detail. Suppose that after a
unitary transformation we bring a Hamiltonian to the form $
H_{diag} + V $ where $V$ contains only terms with at least one
creation operator {\it and} one annihilation operator. That is,
$V$ can be written as $V=a^{\dagger}_i M_{i j} a_j$ where $M$ is
an arbitrary operator and $i,j$ are generic indices. Then,
standard perturbation theory tell us that the correction to an
arbitrary state $|n\rangle$ is given by
\begin{equation}
|\delta n\rangle = \sum_{l\neq n} {
\langle l|V|n\rangle \over (E_l-E_n)}|l\rangle +
\sum_{l,n\neq m}{\langle l|V|m\rangle
\langle m|V|n\rangle \over (E_l-E_m)(E_m-E_n)}|l\rangle +\cdots
\end{equation}
Therefore, if $|n\rangle \equiv |0\rangle$, the vacuum state, it is
annihilated by $V$ and there is no correction to it at any order
in perturbation theory. The ground state is unaffected by $V$.

Now we can proceed to find the unitary transformation of
eq.~(\ref{hr2}). First we split the original Hamiltonian into
four pieces,
\begin{equation}
H=H_1+H_2+V_A + V_B.
\label{hr3}
\end{equation}
Here $H_1$ contains only the modes with momenta less than $\mu$,
$H_2$ is the {\it free} part for all the modes with momenta
greater than $\mu$ and $V_A + V_B$ contains mixing terms and all
non-diagonal high-momentum operators; $V_A$ contains the ``pure''
terms that have only high frequency creation operators or high
frequency annihilation operators, but not both, and $V_B$ the
``impure'' remaining terms (we assume here that $V_A$ and $V_B$
are normal-ordered with respect to the free perturbative vacuum).
Then we break the unitary operator $U$ in a product series,
$U=U_0\; U_1 \cdots U_n \cdots$ and we compute all the terms
successively. The objective of each individual $U_n$ is to
partially diagonalize the Hamiltonian, at a given order in
$\lambda$ (a generic coupling of the theory) and $\mu/\Lambda$.
Let us show how to achieve this.

We proceed iteratively by first diagonalizing the Hamiltonian at
leading order in $\lambda$, up to the desired accuracy in
$\mu/\Lambda$. Consider first a unitary operator $U_0$ written as
$e^{i \Omega_0}$. Now we perform a unitary transformation on
equation (\ref{hr3}), expanding in powers of $\Omega$ (in the
general case $\Omega$ is at least of order $\lambda$ so at a
given order only a finite number of terms are needed):
\begin{eqnarray}
e^{-i\Omega_0}(H_1+H_2+V_A + V_B)e^{i\Omega_0}&\!=&\!
H_1+H_2+V_A + V_B +
i\left[ H_1,\Omega_0\right]+ \nonumber\\
&&\hspace{-1cm} +\, i\left[ H_2,\Omega_0\right]+
i\left[V_A,\Omega_0\right]+i\left[V_B,\Omega_0\right]\cdots
\label{hr4}
\end{eqnarray}
As we have explained above, we want to eliminate the ``pure''
mixing terms $V_A$, while we generate an expansion in
$\mu/\Lambda$. Thus, we impose the following condition on
$\Omega_0$,
\begin{equation}
i\left[ H_2,\Omega_0\right]+V_A=0.
\label{hr5}
\end{equation}

This equation can be solved perturbatively and since commutators
with $H_2$ generate time derivatives of the high frequency fields we
have the desired expansion. Notice that equation (\ref{hr5})
defines $\Omega_0$  up to terms that commute with $H_2$. This
ambiguity corresponds to the redefinition of the low energy
Hamiltonian $H_1$ by a unitary transformation. We will assume
some kind of a ``minimal" scheme - namely that $\Omega_0$ does not
have a part that commutes with $H_2$.

After $\Omega_0$ is chosen to cancel $V_A$ in the effective
Hamiltonian, a new mixing term of order $\lambda$ appears from
$i\left[ H_1,\Omega_0\right]$. However this new term is of a
higher order in $\mu/\Lambda$ and will be eliminated by the next
unitary transformation $U_1 = e^{i\Omega_1}$. Explicitly,
\begin{eqnarray}
e^{-i\Omega_1}e^{-i\Omega_0}(H_1+H_2+V_A + V_B)
e^{i\Omega_0}e^{i\Omega_1} &\!\!\!=&\!\!\! H_1+H_2 +V_B
+\nonumber\\ &&\hspace{-7cm} i\left[ H_1,\Omega_0\right]+
 i\left[ H_2,\Omega_1\right]+
i\left[H_1,\Omega_1\right]+ {\rm second\;order\; terms}\cdots
\label{hr6}
\end{eqnarray}
and we now choose $\Omega_1$ so that
\begin{equation}
i\left[ H_1,\Omega_0\right]+i\left[ H_2,\Omega_1\right] =0.
\label{hr7}
\end{equation}
Using equations (\ref{hr6}) and (\ref{hr7}) we obtain:
\begin{eqnarray}
e^{-i\Omega_1}e^{-i\Omega_0}(H_1+H_2+V_{12})
e^{i\Omega_0}e^{i\Omega_1} &\!\!\!=&\!\!\! H_1+H_2 + V_B +
{i\over 2}\left[V_A,\Omega_0\right]+
\nonumber\\
&&\hspace{-6cm}i\left[V_B,\Omega_0\right]+
i\left[V_B,\Omega_1\right]
+ i\left[H_1,\Omega_1\right]-
{1\over 2}\left[\left[H_1,\Omega_0\right],\Omega_0\right]
-\nonumber\\
&&\hspace{-6cm}
{1\over 2}\left[\left[H_1,\Omega_0\right],\Omega_1\right]
-{1\over 2}\left[\left[H_1,\Omega_1\right],\Omega_1\right]+
\; {\rm higher\;order\; terms}\cdots
\label{e8}
\end{eqnarray}

Now it is easy to deduce the logic of the procedure: the next
step is to introduce $\Omega_2$ in order to cancel
$i\left[H_1,\Omega_1\right]$, which is of higher order in
$\mu/\Lambda$, and continue with the same process until we have
attained the desired accuracy. Notice that by construction
\begin{equation}
\left[ H_2,\Omega_{n+1}\right]=-\left[ H_1,\Omega_n\right]
\label{e9}
\end{equation}
and $H_2\approx\Lambda, H_1\approx\mu$, then
\begin{equation}
\Omega_{n+1}\approx{\mu\over\Lambda}\; \Omega_n ,
\label{e10}
\end{equation}
and any new $\Omega$ is smaller than the previous one by a factor
of $\mu/\Lambda$.

Of course so far we have only eliminated the high momenta degrees
of freedom up to the first order in the coupling constant.
Requiring the absence of the $\lambda^2$ - order mixing terms will
lead to the introduction of a whole new series of unitary
transformations, and the same arguments can be applied to them.

As a final remark, notice that the ``partial" diagonalization at
first order in $\lambda$ defines a correct low-energy effective
Hamiltonian (after projecting onto the high frequency vacuum)
which is valid up to the order $\lambda^3$. The reason is that
since all the relevant terms of order $\lambda$ have been
cancelled, the introduction of the new series of $\Omega$'s of
order $\lambda^2$ will only modify the Hamiltonian at order
$\lambda^4$.

%%%%%%%%%%%%%%%%%%%%%%%%%%%%%%%%%%%%%%%%%%%%%%%%%%%%%%%%%%%%%%%%%%%%%%%%
%%%%%%%%%%%%%%%%%%%%%%%%%%%%%%%%%%%%%%%%%%%%%%%%%%%%%%%%%%%%%%%%%%%%%%%%
%%%%%%%%%%%%%%%%%%%%%%%%%%%   SECTION  II  %%%%%%%%%%%%%%%%%%%%%%%%%%%%%
%%%%%%%%%%%%%%%%%%%%%%%%%%%%%%%%%%%%%%%%%%%%%%%%%%%%%%%%%%%%%%%%%%%%%%%%
%%%%%%%%%%%%%%%%%%%%%%%%%%%%%%%%%%%%%%%%%%%%%%%%%%%%%%%%%%%%%%%%%%%%%%%%

\section{A quantum-mechanical example}

In this section we illustrate the ideas of the previous section
by studying in some detail a quantum mechanical ``toy model"
where the main ingredients of our technique can be presented in a
simpler context.

Consider the following Hamiltonian of two coupled anharmonic
oscillators,
\begin{equation}
H=\frac{{\omega}_1 }{2} \left(P^2 + Q^2\right) +
\frac{{\omega}_2 }{2} \left(p^2 + q^2\right) +
{\lambda}\left( Q + q \right)^4
\label{qm1}
\end{equation}
where $\omega_1<<\omega_2$. Now we want to find the effective
Hamiltonian for $P$ and $Q$ resulting from ``integrating out" the
high frequency modes $p$ and $q$. We choose to compute the
effective Hamiltonian up to to order $\lambda^3$ and
$(\omega_1/\omega_2)^3$.

According to the discussion of the previous section we divide the
Hamiltonian in four pieces as in eq. (\ref{hr3}),
\begin{equation}
H=H_1+H_2+V_A + V_B
\label{qm2}
\end{equation}
where the various terms have the form:
\begin{eqnarray}
H_1&=& \frac{\omega_1}{2} \left(P^2 + Q^2\right) +
{\lambda}\left(\frac{3}{4} + 3\; Q^2 + Q^4\right),\nonumber\\
H_2&=& \omega_2\; a^{\dagger} a ,\nonumber\\
V_A&=& \lambda \left\{ \frac{3}{2} a^2 + \frac{3}{2} a^{\dagger 2}+
\frac{1}{4} a^4 + \frac{1}{4} a^{\dagger 4} +
\sqrt{2}\; Q \left( 3 a + 3 a^{\dagger} +
a^3 + a^{\dagger 3}\right) +\right.\nonumber\\
&&\left. 3\; Q^2 \left( a^2 +  a^{\dagger 2}\right) +
2^{\frac{3}{2}} Q^3 \left( a +  a^{\dagger} \right) \right\},
\nonumber\\
V_B&=&\lambda \left\{ 3 a^{\dagger} a + a^{\dagger} a^3 +
6 a^{\dagger 2} a^2 +  a^{\dagger 3} a +
3\sqrt{2}\; Q \left( a^{\dagger} a^2 + a^{\dagger 2} a
\right) +\right.\nonumber\\
&&\left. 6\; Q^2\; a^{\dagger} a \right\}.
\label{qm3}
\end{eqnarray}
In eq. (\ref{qm3}), $a$ and $a^{\dagger}$ are the annihilation and
the creation operators respectively of the high frequency states,
defined through $a\equiv (p + i q)/\sqrt{2}$ and
$a^{\dagger}\equiv (p - i q)/\sqrt{2}$.

Now we are ready to perform the first unitary transformation
$U_0=e^{i \Omega_0}$. As explained in Section 1, $\Omega_0$
is chosen in order to satisfy equation (\ref{hr5}). Using the
expressions given in eq. (\ref{qm3}) we obtain:
\begin{eqnarray}
\Omega_0&=& i \lambda \frac{1}{\omega_2}
\left\{ \frac{3}{4} ( a^{\dagger 2} - a^2) +
\sqrt{2}\; Q \left( 3 (a^{\dagger} - a) +\frac{1}{3}
( a^{\dagger 3} - a^3) \right) +\right.\nonumber\\
&&\left. \frac{3}{2}\; Q^2 \left(a^{\dagger 2}-a^2\right) +
2^{\frac{3}{2}} Q^3 \left( a^{\dagger} -a  \right)
+ \frac{1}{16} (a^{\dagger 4} - a^4) \right\}.
\label{qm4}
\end{eqnarray}

After performing the unitary transformation with the above
expression for $\Omega_0$, we obtain the leading order
approximation of the diagonalization of the whole Hamiltonian: we
have eliminated the non-diagonal ``pure" terms $V_A$ at the
expense of creating new ones in $[H_1,\Omega_0]$. However those
new terms are of order of $\omega_1/\omega_2$ as can be seen
clearly from eqs. (\ref{qm3}) and (\ref{qm4}).

The following step is then to cancel this newly created
non-diagonal terms with the next unitary transformation $U_1=e^{i
\Omega_1}$ where $\Omega_1$ satisfies equation (\ref{hr7}). A
straightforward calculation gives
\begin{eqnarray}
\Omega_1&=& i \lambda \frac{\omega_1}{\omega_2^2}
\left\{ \frac{3}{4} ( a^{\dagger 2} + a^2) +
i \sqrt{2}\; P \left( 3 (a^{\dagger} + a) +\frac{1}{9}
( a^{\dagger 3} + a^3) \right) +\right.\nonumber\\
&&\left. i \frac{3}{2}\; Q P \left(a^{\dagger 2} + a^2\right) -
i 6 \sqrt{2} Q^2 P\left( a^{\dagger} + a  \right)\right\}.
\label{qm5}
\end{eqnarray}
Finally, to reach the desired accuracy in
$\frac{\omega_1}{\omega_2}$ we need also $\Omega_2$, defined
through the relation $[H1,\Omega_1] + [H_2,\Omega_2]=0$. We find
\begin{eqnarray}
\Omega_2&=& - \lambda \frac{\omega_1^2}{\omega_2^3}
\left\{ 12 ( a^{\dagger} - a ) -
i \sqrt{2}\; Q \left( 3 (a^{\dagger} - a) -\frac{1}{27}
( a^{\dagger 3} - a^3) \right) -\right.\nonumber\\
&&\hspace{-2cm}\left. i\; \frac{3}{4}\; (P^2 + Q^2)
\left(a^{\dagger 2} - a^2\right) +
i\; 12 \sqrt{2}\; Q P^2\left( a^{\dagger} -  a  \right)
i\; 6 \sqrt{2}\; Q^3 \left( a^{\dagger} -  a  \right)
\right\}.
\label{qm6}
\end{eqnarray}

Now we are ready to compute the ``low energy" effective
Hamiltonian. Recalling the discussion of the previous section, we
have the following expression for the effective Hamiltonian, up to
the order $\lambda^3$ and $1/\omega_2^3$:
\begin{eqnarray}
H_{eff} &\hspace{-.1cm}=&\hspace{-.1cm} \langle 0| \left\{ H_1 +
H_2 + V_B + \frac{i}{2} [V_A,\Omega_0] + i [V_B,\Omega_0] +  i
[V_B,\Omega_1] + i [V_B,\Omega_2] - \right.\nonumber\\
&&\hspace{-.35cm} \frac{1}{2} [[H_1, \Omega_0],\Omega_0] -
\frac{1}{2} [[H_1, \Omega_0],\Omega_1] - \frac{1}{3} [[V_A,
\Omega_0],\Omega_0] - \frac{1}{2} [[V_A, \Omega_0],\Omega_1]
-\nonumber\\
&&\hspace{-.35cm}\left. \frac{1}{2} [[V_B, \Omega_0],\Omega_0]-
[[V_B, \Omega_0],\Omega_1] - \frac{i}{6} [[[H_1,
\Omega_0],\Omega_0],\Omega_0] \right\} |0 \rangle . \label{qm7}
\end{eqnarray}
Here $|0\rangle$ is the vacuum associated to the operators $a$
and $a^{\dagger}$ leaving the $P$ and $Q$ operators untouched.
For the sake of completeness we have written the whole expression
derived from the rules described in Section 1. However, some of
the terms vanish upon projection.

Finally after a long but straightforward computation we have the
desired low frequency Hamiltonian  :
\begin{eqnarray}
H_{eff} &\hspace{-.1cm}=&\hspace{-.1cm} \frac{\omega_1}{2}
\left(P^2 + Q^2\right) + {\lambda}\left(\frac{3}{4} + 3\; Q^2 +
Q^4\right) + \lambda^2 \left( -\frac{21}{8} \frac{1}{\omega_2} +
\frac{29}{3} \frac{\omega_1}{\omega_2^2} +\right. \nonumber\\
&&\hspace{-.7cm} \frac{153}{4} \frac{\omega_1^2}{\omega_2^3}
+\left(-31 \frac{1}{\omega_2} + 45 \frac{\omega_1}{\omega2^2} +
144 \frac{\omega_1^2}{\omega_2^3}\right) Q^2 - \frac{166}{9}
\frac{\omega_1^2}{\omega_2^3}\; P^2 +\nonumber\\
&&\hspace{-.7cm}
162\; i\; \frac{\omega_1^2}{\omega_2^3}\; Q\; P - 81
\frac{\omega_1^2}{\omega_2^3}\; Q^2 P^2 + 288\; i\;
\frac{\omega_1^2}{\omega_2^3}\; Q^3 P + \left(36
\frac{\omega_1}{\omega_2^2} - \frac{33}{\omega_2} \right) Q^4
+\nonumber\\
&&\hspace{-.7cm} \left. 72
\frac{\omega_1^2}{\omega_2^3}\; Q^4 P^2 - \frac{8}{\omega_2}\;
Q^6\right)+ \lambda^3 \left( \frac{333}{16} \frac{1}{\omega_2^2}
- 178 \frac{\omega_1}{\omega_2^3} +
\left(\frac{168}{4}\frac{1}{\omega_2^2} -\right.
\right.\nonumber\\
&&\hspace{-.7cm} \left. \frac{3653}{2}
\frac{\omega_1}{\omega_2^3} \right) Q^2 +
\left(\frac{888}{\omega_2^2} - 3132 \frac{\omega_1}{\omega_2^3}
\right) Q^4 +  \left(\frac{534}{\omega_2^2} - 1224
\frac{\omega_1}{\omega_2^2} \right) Q^6 +\nonumber\\
&&\hspace{-.7cm} \left. \frac{96}{\omega_2^2}\; Q^8 \right) +
{\it O} (\lambda^4,1/\omega_2^4). \label{qm8}
\end{eqnarray}

The skeptical reader can verify, using standard
Rayleigh-Schr\"odinger perturbation theory, that the spectrum of
the effective Hamiltonian (\ref{qm8}) is the same as the low
energy spectrum ({\it i.e.,} the part of the spectrum that
remains finite if $\omega_2 \to \infty$) of the original
Hamiltonian (\ref{qm1}).

%%%%%%%%%%%%%%%%%%%%%%%%%%%%%%%%%%%%%%%%%%%%%%%%%%%%%%%%%%%%%%%%%%%%%%%%
%%%%%%%%%%%%%%%%%%%%%%%%%%%%%%%%%%%%%%%%%%%%%%%%%%%%%%%%%%%%%%%%%%%%%%%%
%%%%%%%%%%%%%%%%%%%%%%%%%%%   SECTION  III %%%%%%%%%%%%%%%%%%%%%%%%%%%%%
%%%%%%%%%%%%%%%%%%%%%%%%%%%%%%%%%%%%%%%%%%%%%%%%%%%%%%%%%%%%%%%%%%%%%%%%
%%%%%%%%%%%%%%%%%%%%%%%%%%%%%%%%%%%%%%%%%%%%%%%%%%%%%%%%%%%%%%%%%%%%%%%%

\section{Scalar Field Theory }

In this section we apply the same formalism to the case of a
scalar field with the quartic self-interaction, $\lambda\;
\phi^4$. We will determine the effective Hamiltonian $H$ up to
the two-loop order.

Before proceeding to the loop calculation we have to explain how
renormalization is performed in our formalism. As was shown in
Section 1, by projecting to the vacuum state for the ``high''
momentum modes we obtain the effective Hamiltonian $H_{eff}$.
Since our system is supposed to have had an ultraviolet cut-off
$\Lambda$ from the very beginning, $H_{eff}$ will explicitly
depend on this UV scale. The renormalization procedure consists
of modifying the original Hamiltonian $H$ by introducing
renormalization $Z$-factors that depend on the UV cut-off
$\Lambda$ and some arbitrary ``renormalization scale'' $M$. Each
of the $Z$'s depends on $\Lambda$ and $M$ in such a way that the
effective Hamiltonian obtained from it does not depend on
$\Lambda$; in fact it has to look exactly like the original one
except that all $\Lambda$'s should be replaced by $\mu$ - the
scale down to which we are integrating out.

Let us start from the ``bare'' Hamiltonian
\begin{equation}
H=\int d^3x \left({1\over 2} \Pi^2_\Phi(x) + {1\over 2}
\Phi(x)\left( -\grad^2+ m^2 \right)\Phi(x)+ \lambda
\Phi^4(x)\right) , \label{g2}
\end{equation}
 and introduce a $Z$ factor for each composite operator,
\begin{equation}
H=\int \left( \frac{Z_\pi}{2} \Pi^2_\Phi(x) + \frac{Z_\phi}{2}
\Phi(x)\left( -\grad^2\right)\Phi(x) + Z_m Z_\phi m^2 \Phi^2(x)+
\lambda Z_\lambda Z^2_\phi \Phi^4(x)\right) \label{g21}
\end{equation}
Each of the $Z$-factors has a perturbative expansion in
$\lambda$, where we are assuming that $\lambda$ has been defined
by some renormalization prescription at the renormalization scale
$M$ ($\lambda$ is the ``renormalized" coupling in the language of
the standard renormalization group). Generically:
\begin{equation}
Z = 1 + f_1(\Lambda) \lambda + f_2(\Lambda) \lambda^2 +
f_3(\Lambda) \lambda^3 + \cdots . \label{zeta}
\end{equation}
The functions $f_n$ will be chosen order-by-order from the
requirement that after integration of the modes from $\mu$ to
$\Lambda$ all the corrections sum up in such a way that
$Z(\Lambda)\rightarrow Z(\mu)$. When doing the one-loop
corrections one can therefore assume that all the $Z$'s are
initially 1 and choose the corresponding $f$'s from the condition
that high-cutoff dependence be cancelled after computing the
$H_{eff}$ to one loop. For the second loop we use the one-loop
$Z$-corrected Hamiltonian and determine $f_n$ to the next order in
$\lambda$ by the same procedure.

According to our philosophy we have to identify the purely ``low"
part $H_1$, the free ``high" momentum part $H_2$ and the
interaction terms $V_A$ and $V_B$. First we split the original
field $\Phi(x)$ into its low and high frequency components:
\begin{eqnarray}
\Phi(x)&=& \phi(x) + \chi(x) ,\nonumber\\ \phi(x)&=&\sum_{k<\mu}
\Phi_k e^{ikx}, \ \ \ \ \chi(x)=\sum_{\mu<k<\Lambda} \Phi_k
e^{ikx} .\label{g3}
\end{eqnarray}
By virtue of the equations (\ref{g3}), $\;\int \phi(x)\chi(x) =
0 $ and the original Hamiltonian (\ref{g2}) can be rewritten as
sum on the four pieces $H=H_1+H_2+V_A+V_B$:
\begin{eqnarray}
H_1&=&\int\left( \frac{1}{2} \pi^2_\phi(x) + {1\over 2}
\phi(x)\left(-\grad^2+ m^2 \right)\phi(x)+ \lambda
\phi^4(x)\right) ,\label{g4}\\
H_2&=&\int \left( {1\over 2} {\pi^2_\chi(x)} + {1\over 2}
\chi(x)\left( -\grad^2+ m^2 \right)\chi(x)\right) ,\label{g5}\\
V_A+V_B &=& \lambda \int \left( \chi^4(x)+  4\phi\chi^3 +
6\phi^2\chi^2 + 4\phi^3\chi \right) .\label{g6}
\end{eqnarray}
In order to separate $V_A$ and $V_B$, according to the discussion
of Section 1, we introduce the second-quantized form of the
high-momentum modes $\chi$,
\begin{equation}
\chi(x)=\sum_{\mu<k<\Lambda} {1\over \sqrt{2\omega_k}}\left(a_k
e^{ikx} + a^\dagger_k e^{-ikx}\right) \label{g7}
\end{equation}
Here $\omega_k=\sqrt{k^2+m^2}$ and can be taken $\omega_k\approx
|k|$ for $\mu >> m$. The operators $a_k$ and $a_k^\dagger$
satisfy the standard commutation relations
$[a_k,a^\dagger_k]=\delta_{kp}$. Upon substituting this
definition in the expressions (\ref{g5}) and (\ref{g6}) and
normal-ordering the result with respect to ``high'' modes we can
identify all the terms that contain only creation or only
annihilation operators; these terms form $V_A$. The rest of the
interaction part forms $V_B$.
\begin{eqnarray}
H_2 &=& \sum_{\mu<k<\Lambda} \omega_k a^\dagger_k a_k ,
\label{g8}\\
V_A &=& \lambda\sum\left\{ {e^{i(k+p+q+r)x} a_k a_p a_q a_r \over 4
\sqrt{\omega_k \omega_p \omega_q \omega_r}} + {2\;\phi(x)
e^{i(k+p+q)x} a_k a_p a_q \over \sqrt{2\;\omega_k \omega_p
\omega_q}} + \right. \nonumber\\ &&\hspace{-.8cm}\left.
{3\;\phi^2(x)\; e^{i(k+p)x} a_k a_p\over \sqrt{\omega_k
\omega_p}}  + {4\;\phi^3(x) e^{ikx} a_k\over\sqrt{2\;\omega_k}}
+ \frac{3}{2} \frac{1}{\omega_p}\; \frac{a_k a_{-k}}{\omega_k}
+ {\rm h.\; c.}\right\} , \label{g9}\\
V_B &=& \lambda\sum\left\{ {e^{i(p+q+r-k)x} a^\dagger_k a_p a_q a_r
\over \sqrt{\omega_k \omega_p \omega_q \omega_r} }
+{3\;e^{i(q+r-k-p)x} a^\dagger_k a^\dagger_p a_q a_r \over
\sqrt{\omega_k \omega_p \omega_q \omega_r} } \right. \nonumber \\
&&\hspace{0.0cm}\left.+{6\;\phi(x) e^{i(p+q-k)x} a^\dagger_k a_p
a_q \over \sqrt{2\;\omega_k \omega_p \omega_q}}+ {\rm h.\; c.}\right\}
\end{eqnarray}
Notice that already at this stage we have some contributions to
the effective Hamiltonian due to the normal-ordering of the terms
of type $\phi^2(x)a_k a^\dagger_p$; this term  explicitly
depends on the UV cut-off $\Lambda$ and can be included into
$H_1$,
\begin{equation}
\delta H_1=6\lambda \sum_{\mu<k<\Lambda}{1\over 2\omega_k}\; \int
\phi^2(x) \approx 3\;\lambda \frac{\Lambda^2 - \mu^2}{4 \pi^2}\;
\int \phi^2(x), \label{mass}
\end{equation}
where we have made the standard replacement $\sum_k \to \int {d^3
k\over (2 \pi)^3}$. Equation (\ref{mass}) is, of course, the
standard ``tadpole'', one-loop  mass renormalization.

In order to pick up the only other one-loop contribution, the
coupling constant renormalization, we have to follow the general
procedure of Section I and determine $\Omega_0$ for the first
unitary transformation. Using equation (\ref{hr5}) we deduce
\begin{eqnarray}
\Omega_0 &=& (-i) \lambda\sum \left\{ {4\;\phi^3(x) e^{ikx}
a_k\over\sqrt{2\omega_k}\;\omega_k} + {3\;\phi^2(x) e^{i(k+p)x}
a_k a_p\over \sqrt{\omega_k\omega_p}(\omega_k+\omega_p)} +
\frac{3}{4} \frac{1}{\omega_p}\; \frac{a_k a_{-k}}{\omega^2_k} +
\right.\nonumber\\
&&\hspace{-1.8cm}\left.{2\phi(x) e^{i(k+p+q)x}
a_k a_p a_q \over \sqrt{2\;\omega_k \omega_p
\omega_q}(\omega_k+\omega_p+\omega_q)}+ {e^{i(k+p+q+r)x} a_k a_p
a_q a_r \over 4\;\sqrt{\omega_k \omega_p \omega_q \omega_r}
(\omega_k+\omega_p+\omega_q+\omega_r)} - {\rm h.c.} \right\}
\label{Omega0}
\end{eqnarray}

To find the correction to the Hamiltonian we expand the unitary
transformations and project onto the high energy vacuum. The
contributions to $H_{eff}$ are exactly the same as in equation
(\ref{qm7}).

Now we have to determine the potentially divergent contributions
(if $\Lambda \to \infty$) that will emerge after projection to
the vacuum state. In doing so the following naive power-counting
rule is useful:

a) Each $\Omega_n$ brings $k^{(n+1)}$ (momentum) to the
denominator.

b) Each contraction contributes with $k^{-1}$.

c) For each loop integration we have a contribution of $k^3$.

Therefore, each term for which the total degree of divergence of
a term is greater than or equal to zero, has to be taken into
account. Of course, there may be overlapping divergences in
computing the proper expressions, but those have to be studied
individually, term by term.

By inspection of equation (\ref{qm7}), is easy to see that at
this order there are no contributions to the terms $\Pi^2$ and
$\Phi( -\grad^2)\Phi$ ({\it i. e.,} $Z_\pi=Z_\phi=1$) and the
only other one-loop term is the coupling constant
renormalization. This contribution comes from the commutator
$[V_A, \Omega_0]$ when contracting twice the high energy
fields. The rest of the terms are not important at this stage:
the commutators $[V_B, \Omega_0]$ and $[H_1,\Omega_0]$ are zero
when projected onto the vacuum. After projecting onto the high
frequency vacuum we have
\begin{eqnarray}
\delta {H_1}|^{\lambda^2}_{\phi^4}&=& -18 \lambda^2\int
\phi(x)\phi(y)\; {dp\;dq\over (2\pi)^6}\; {1\over \omega_k
\omega_p(\omega_k+\omega_p)}\; e^{i(p-q)(x-y)} \nonumber\\ &=&
-{9\lambda^2 \over 2\pi^2}\; \ln\left({\Lambda\over\mu}\right)\;
\left[\int\phi^4(x)\;d^3x\right] .
\label{lmbd}
\end{eqnarray}
Expressions (\ref{mass}) and (\ref{lmbd}) give the two
contributions to $H_{eff}$ at one-loop. From them we can deduce
the renormalization factors $Z_m$ and $Z_{\lambda}$:
\begin{equation}
Z_m=1 - 3\lambda{\left(\Lambda^2- M^2\right)\over 2\pi^2}\ ,\ \ \
Z_{\lambda}=1 + {9\lambda^2 \over 2\pi^2}\;
\ln\left({\Lambda\over M}\right) .\label{one-loop}
\end{equation}

This finishes the one-loop renormalization procedure. (In refs.
\cite{Nair-Minic,sz} there is an extra one-loop counterterm of order
$\lambda$ due to the definition of the ``effective" wave
functional in the Schr\"odinger representation). Before going to
the next order in coupling, we should notice that there is a
two-loop contribution coming from the same term $[V_A,\Omega_0]$
in (\ref{qm7}), when contracting the high energy fields three
times. Its leading divergence is quadratic and  it has a
logarithmic subleading divergent part that gives rise to the
two-loop wave-function renormalization,
\begin{eqnarray}
\delta {H_1}|^{\lambda^2}_{\phi^2} &\!\!=&\!\! -{12\lambda^2\over
(2\pi)^9}\left\{\int\;d^3r\phi(r)\phi(-r)\right\}
\int{d^3p\;d^3q\over \omega_p \omega_q \omega_{r-p-q} ( \omega_p
+  \omega_q + \omega_{r-p-q})}\nonumber\\
&&\hspace{-1.6cm}=
-{3\lambda^2\over \pi^4}\left(2\;\ln 2 - 1\right) \Lambda^2 \int
{1\over 2} \phi^2(x)\;d^3x + {3\over 8} {\lambda^2\over\pi^4}
\ln\left({\Lambda\over\mu}\right) \int {1\over 2}
\left(\grad\phi(x)\right)^2 \label{g11}
\end{eqnarray}

Extra contributions to $H_{eff}$ at two-loops come from the terms
\begin{eqnarray}
\delta H = \langle 0|\left( -{1\over 2}
\left[\left[H_1,\Omega_0\right],\Omega_1\right] + i
\left[V_B,\Omega_1\right] -{1\over 2}
\left[\left[H_1,\Omega_1\right], \Omega_1\right]\right) |0\rangle
.
\end{eqnarray}
However using the power counting rules described above we can see that
the only divergent contribution comes from the first term.
Following equation (\ref{hr7}) we determine the operator $\Omega_1$:
\begin{eqnarray}
\Omega_1 &=& (-i)\lambda\sum \left\{ {4\left[H_1,\phi^3(x)\right]
e^{ikx} a_k\over \sqrt{2\omega_k}\omega_k^2} +
{3\;\left[H_1,\phi^2(x)\right] e^{i(k+p)x} a_k a_p\over
\sqrt{2\omega_k\omega_p} (\omega_k+\omega_p)^2}+
\right.\nonumber\\
&&\left.\hspace{-2.2cm}
{2\left[H_1,\phi(x)\right] e^{i(k+p+q)x} a_k a_p a_q \over
\sqrt{2\omega_k \omega_p \omega_q}(\omega_k+
\omega_p+\omega_q)^2}+ {e^{i(k+p+q+r)x} a_k a_p a_q a_r \over 4
\sqrt{\omega_k \omega_p \omega_q \omega_r} (\omega_k+
\omega_p+\omega_q+\omega_r)^2} - {\rm h.c.}\right\} \label{Omega1}
\end{eqnarray}
and after evaluating a momentum integral similar to the one in
eq.~(\ref{g11}) we finally get
\begin{equation}
\delta H|_{\pi^2}^{\lambda^2}=- {3\over 8} {\lambda^2\over\pi^4}
\ln\left({\Lambda\over\mu}\right) \left\{\int {1\over 2}
\pi^2_\phi\; d^3x\right\} \label{g12}
\end{equation}
It is important to point out that the corrections to $(\grad
\phi)^2$, eq.~(\ref{g11}), and $\pi^2$  eq.~(\ref{g12}), have
equal magnitude and opposite sign as it must be, due to the
Lorentz covariance of the theory. In fact, since $\pi_\phi$ is
represented by ${\delta\over\delta\phi}$, then if $\phi^2$ gets
corrected by $Z$ then $\pi_\phi^2$ should change by ${1\over Z}$,
so that the equal time commutator is preserved. Now we can define
the two-loop wave-function renormalization factors $Z_\phi$ and
$Z_\pi$ as
\begin{eqnarray}
Z_\phi &=& 1 - {3\over 8} {\lambda^2\over\pi^4}
\ln\left({\Lambda\over\mu}\right) ,\label{wavephi}\\
 Z_\pi &=&
{1\over Z_\phi} = 1+ {3\over 8}
{\lambda^2\over\pi^4}\ln\left({\Lambda\over\mu}\right).
\label{wavepi}
\end{eqnarray}

Finally there is one more two-loop contribution that comes at the
order $\lambda^3$ and renormalizes the $\phi^4(x)$ term. It comes
from the terms
\begin{eqnarray}
\delta H= \langle 0|\left( -{1\over 3}
\left[\left[V_A,\Omega_0\right],\Omega_0\right] -{1\over 2}
\left[\left[V_B,\Omega_0\right],\Omega_0\right] \right) |0\rangle
\end{eqnarray}
After a somewhat tedious computation we get
\begin{equation}
\delta H^{(2-loop)}_{\phi^4}= {27\over 2}
{1\over\pi^4}\lambda^3\left\{\ln\left({\Lambda\over\mu}\right)+
\left[\ln\left({\Lambda\over\mu}\right)\right]^2\right\} \int
\phi^4(x)\;d^3x .
\end{equation}

To compute the entire two-loop correction we have to add to this
result the contribution of the one-loop counterterms ,{\it i.e.,}
the one-loop terms with the order $\lambda$ contributions to the
$Z$-factors. Thus, we can finally deduce the value of the
$Z_\lambda$ factor at two-loops,
\begin{equation}
Z_\lambda = 1 + \frac{9}{2 \pi^2} \log(\Lambda/M)\;
\lambda^2 + \left( \frac{81}{4 \pi^4} (\log (\Lambda/M))^2 -
\frac{51}{4 \pi^4} \log(\Lambda/M)\right)\; \lambda^3.
\label{z2loop}
\end{equation}
{}From equations (\ref{wavephi}) and (\ref{z2loop}) we get the
correct two-loop $\beta$-function of the theory,
\begin{eqnarray}
\beta(\lambda)&=&\left. \frac{\partial \lambda}{\partial \log M }
\right|_{\Lambda}= -\left((\partial_\lambda ( \lambda\;
Z_\lambda)|_{\Lambda, M} \right)^{-1} \lambda \left.
\frac{\partial Z_\lambda}{\log M }\right|_{\lambda,\Lambda}
\nonumber\\ &=& \frac{9}{2 \pi^2} \lambda^2 - \frac{51}{4 \pi^4}
\lambda^3 \;, \label{beta}
\end{eqnarray}
and the correct two-loop anomalous dimension $\gamma$:
\begin{equation}
\gamma(\lambda) = \frac{1}{2} \left.\frac{\partial \log
Z_\phi}{\partial \log M } \right|_{\Lambda}= \frac{3}{16 \pi^4}
\lambda^2 \label{gamma}.
\end{equation}

In summary, we have described in this letter a novel perturbative
technique of renormalization in the Hamiltonian (Schr\"odinger)
formalism. We have showed that this method successfully gives the
two-loop renormalized Hamiltonian for a scalar Field Theory with
a quartic potential. Moreover, in a previous work \cite{AM} we
have shown that this technique also gives a consistent one-loop
renormalization for the Quantum Electrodynamics and the
Yang-Mills Hamiltonians where results similar to those from
Coulomb gauge covariant calculations were found.

\section{Acknowledgements}

We wish to thank V.~P.~Nair and R.~Ray for helpful discussions
and a critical reading of the manuscript. G.A. was partially
supported by CUNY RF grant 6684591433. E.F.M. was partially
supported by CUNY Collaborative Incentive Grant 991999.

\end{document}